\documentclass{article}

\usepackage{arxiv}

\usepackage[utf8]{inputenc} % allow utf-8 input
\usepackage[T1]{fontenc}    % use 8-bit T1 fonts
\usepackage{hyperref}       % hyperlinks
\usepackage{url}            % simple URL typesetting
\usepackage{booktabs}       % professional-quality tables
\usepackage{amsfonts}       % blackboard math symbols
\usepackage{amsmath}
\usepackage{amssymb}
\usepackage{bbold}          % for indicator 1
\usepackage{nicefrac}       % compact symbols for 1/2, etc.
\usepackage{microtype}      % microtypography
\usepackage{graphicx}
\usepackage{natbib}
\usepackage{doi}
\usepackage[usenames,dvipsnames]{color}
\usepackage{xcolor}

\usepackage{listings}

\colorlet{lightorange}{orange!05}
\title{Development and performance of npd for the evaluation of models with ordinal data}
\date{May 4th, 2026}	% Here you can change the date presented in the paper title
\author{ \hspace{1mm}Marc Cerou \\
Univ Rennes, Inserm, EHESP, \\
Irset (UMR S\_1085)\\
F-35000 Rennes \\
    \& Institut de Recherches Internationales Servier \\ 
    F-92150 Suresnes \\
    France \\
	\And
	\hspace{1mm}Marylore Chenel \\
    Institut de Recherches Internationales Servier \\ 
    F-92150 Suresnes, France \\
    \& Pharmatheus AB \\ 
    Uppsala, Sweden \\
	\And
\href{https://orcid.org/0000-0002-9150-9886}{\includegraphics[scale=0.06]{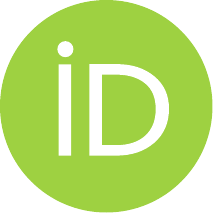}\hspace{1mm}Emmanuelle Comets}\thanks{Corresponding author} \\
Univ Rennes, Inserm, EHESP, \\
Irset (UMR S\_1085)\\
F-35000 Rennes \\
\& Université Paris Cité \& \\
Université Sorbonne Paris Nord\\ 
Inserm, IAME, F-75018 Paris, France\\
	\texttt{emmanuelle.comets@inserm.fr} \\
    }

\renewcommand{\headeright}{npde to evaluate joint models in NLMEM}
\renewcommand{\shorttitle}{{\sf arXiv} preprint}

\hypersetup{
pdftitle={npdeDiscreteModels},
pdfauthor={Marc Cerou, Marylore Chenel,Emmanuelle Comets},
pdfkeywords={Non-linear mixed effect models, Model evaluation, npd, Residuals, Toe-nail infection, longitudinal data, ordinal data, count data, binary data},
}

\def\pd{{\textsf{pd}}}

\def\npd{{\textsf{npd}}}
\def\npde{{\textsf{npde}}}

\begin{document}
\newcommand{\un}{\mathds{1}}
\maketitle

\paragraph{Corresponding author:} Emmanuelle Comets \\
\begin{tabular}{p{3.2cm} l}
&INSERM - UMR1137 \\
&UFR de Médecine Site Bichat \\
&16 rue Henri Huchard \\
& 75018 Paris, France \\
&Tel: (33) 6 25 82 49 50\\
&\texttt{emmanuelle.comets@inserm.fr} \\
\end{tabular}

\newpage
\begin{abstract} % ToDo: 1557/1920 characters
\textbf{Purpose:} Normalised prediction distribution errors (npde) are used to graphically and statistically evaluate continuous responses in non-linear mixed effect models. Here, our aim was to extend npde for categorical data and to evaluate their performance. We applied our approach to a real case-study describing the evolution of severe onychomycosis (toenail infection) in a trial comparing two treatment groups.

\textbf{Methods:} Let V denote a dataset with categorical observations. The null hypothesis $H_0$ is that observations in V can be described by a model M. Residuals called npde can be adapted to categorical observations using jittering techniques. Their theoretical standard normal distribution can be evaluated through the Kolmogorov-Smirnov test. We evaluated the performance in terms of power through a simulation and compared it to a Chi-square. We illustrated the test and graphs on a real case-study. 

\textbf{Results:} npd were able to detect misspecifications in the structural model and model parameter's value. As expected, the power to detect model misspecifications increased both with the difference in the shape of the probability, and with the sample size. Chi-square test performed better but npd could be readily applied in all type of design. Based on the toe-nail data, graphs reveal a huge discrepancy of the base model, and a good adequation for the best model we found.    

\textbf{Conclusions:} npde can be extended to categorical data, particularly in clinical settings with unbalanced design and graphs can be useful to evaluate the model as well as the covariate effect.

\end{abstract}

% keywords can be removed

\newpage
\section{Introduction}
%{\etalchange\citet*{nguyen_extension_2012}} are cited this way.
%\citep{nguyen_extension_2012}

\hskip 18pt Clinical trials typically follow several outcomes of interest over a period of time, providing information on the evolution of a disease leading to the primary endpoint used for the statistical decision. These outcomes may be continuous, such as biomarkers or drug concentrations, but are often ordered categorical variables, such as %the combined ACR scale measuring change in rheumatoid arthritis symptoms, 
the pain scores measuring the quality of analgesia \citep{sheiner_new_1994}, questionnaire-based depression scales, or scores used to assess cognitive abilities in Alzheimer's disease \citep{ito_exposure-response_2008}.%, or counts of episodes of nausea to measure side-effects of drugs. 
Moreover, some continuous biomarkers may be discretised for clinical interpretation and reported as ordered categorical variables (e.g. in oncology, where neutrophile count are categorised in grades 1 to 4). 

% Refondre avec l'intro et enlever d'ici
Since the 1970s, non-linear mixed-effect models have been increasingly used, mainly with continuous data. Following the evolution of disease over time maximises the information and allows a better characterisation of treatment effect. This is especially important for chronic disease or with a long-term effect where the loss to follow-up may impact standard statistical inferences as shown by \cite{ibrahim_basic_2010}. In recent years with the emergence of item response theory (IRT) within the pharmacometrics world thanks to the work of \cite{ueckert_improved_2014}, mixed effects models with discrete data have been used extensively to improve disease information and decision making even with complex data such as scores combining different questionnaires. It allows the analysis of multifactorial discrete data and aimed at describing disease dimension, which can be unique (HAMD scale used to measure depression in ~\cite{cerou_IRTdepression_2019}) or multiple (UPDRS scale in Parkinson in \cite{buatois_item_2017}). This prompted the development and evaluation of new estimation algorithms % \cite{savic_implementation_2010} 
and many innovative approaches to model jointly disease progression and treatment effect both in short and long-term studies, such as the work of \cite{gottipati_modeling_2017}. The evaluation of these models has not however been studied as extensively, in part because analyses often use simple models to focus instead on selecting covariate effects through statistical criteria, and in part because of the more limited information available in discrete observations to produce meaningful diagnostic graphs. 

Model evaluation is an important step in pharmacometrics and consists in checking whether a given model can describe the data and in evaluating the underlying model assumptions. As recommended in \cite{nguyen_model_2017}, model evaluation is required both in the model building, which is the process of model development to achieve defined objectives, and for model qualification, which is the assessment of the model performance with respect to the objectives of the analysis. %This is of concern to avoid waste in human resources, time, and budget by making analyses based on poor decisions. 
Model evaluation is therefore recommended by regulatory agencies (\cite{food_and_drug_administration_guidance_2003}, \cite{agency_guideline_2007}) and Model-Informed drug development (MIDD) is an integral part of the drug development process with its regulatory guidelines and its own good practices as describes in~\cite{marshall2016good}. For continuous data, numerous tools for model evaluation have been developed, and a white paper by %the model evaluation group of the ISoP committee 
\cite{nguyen_model_2017} described the different numerical and graphical diagnostics and when we need to use them. Recommended methods include Visual Predictive Check (VPC) as described by \cite{holford_visual_2005}, normalised prediction distribution (npd) presented by \cite{mentre_prediction_2006} for graphical evaluation and normalised prediction distribution error (npde) showed by \cite{brendel_metrics_2006}  for numerical evaluation as a gold standard. 

For ordinal data however, % most of the work has focused on model selection and inference and the issue of appropriate diagnostic tools has seldom been investigated. %Zhang and al. \cite{zhang_statistical_2011} showed recently the lack of effective diagnostic tools to check the validity of model assumptions. 
some effort has been done to develop residuals for regression using ordinal variables %by Liu et al. (2009) \cite{liu_graphical_2009}, Li and Shepherd (2012) \cite{li_new_2012} and 
like in \cite{liu_residuals_2018}. However, these residuals are adapted to more classical regression ordinal models and do not extend readily to repeated data modelled with random effects. To date, for ordinal data, only simulation-based diagnostics based on VPC can be carried out easily. For this type of response, less traditional VPC are created, which characterise the proportion of each category over time and compare it to the predicted median and prediction interval of their probability. This quickly becomes difficult to synthetise when the data has many categories, or when the VPC need to be stratified according to covariates of interest. %This can be likened to the necessity for stratification in VPC for continuous data when covariates enter the model or when the design is non-homogenous, prompting the development of extensions to VPC such as the pcVPC as discussed in \cite{nguyen_model_2017}. 
By contrast, $\npd$ and $\npde$ provide global metrics to evaluate the model, and \cite{brendel_evaluation_2010} proposed to stratified them in groups of interest . It would therefore be interesting to extent these metrics to models with ordinal responses.

In this work, we propose to define and evaluate $\npd$ for categorical data and adjust the associated test by simulation to account on correlation. %Due to the correlation between repeated observations in an individual, the resulting $\npd$ are not independent. In model with continuous data assuming a Gaussian distribution for the observations, $\npde$ can be computed by decorrelating the observations and the predictions before computing the $\pd$, which allows the test based on the resulting $\npde$ to maintain an appropriate type I error \cite{brendel_metrics_2006}. The decorrelation step however relies on the notion that the joint distribution of the observations in the continuous case is approximately multivariate Gaussian, which is no longer true in models with ordinal data. An alternative approach we propose to use here is to correct the p-value test on the $\npd$ by simulating the distribution of the test statistic to take into account the impact of the correlation. 
In Section~\ref{sec:material}, we describe the construction of the $\npd$ and how we propose to correct the test statistic. We then describe a simulation study which was performed to evaluate the performance of this test in terms of type I error and power to detect different types of model misspecifications. We use the setting proposed by \cite{Seurat} in their work on optimal design for studies with binary outcomes. We also apply our approach to a real case-study describing the evolution of severe onychomycosis (toenail infection) in a trial comparing two treatment groups. The results from the simulation study and the application to real data are given in Section~\ref{sec:results}. Finally in the Discussion, Section~\ref{sec:discussion}, we summarise the main findings and discuss practical implementation.

\section{Models and methods} \label{sec:material}

\subsection{Statistical model for categorical data}

\hskip 18pt Let $Y$ denote the response and assume its values are in a fixed and finite set of categories ${c_1,\dots,c_K}$. Let $Y_{ij}, (1\leq i \leq N, 1\leq j \leq n_i)$ be the observations of the outcome in subject $i$ at times $t_{ij}$.

We denote $\pi_{ij}(c_k|\theta_i, t_{ij})$ the individual probability for category $k$ at time $t_{ij}$ in subject $i$, which depends on the individual parameters $\theta_i$.
%\begin{equation}
%    \begin{tabular}{lr}
%        $\pi_{ij}(c_k|\theta_i, t_{ij}) = Pr(Y_{ij}=c_k|\theta_i, t_{ij})$, & $k=1,\dots,K$.  \\
%    \end{tabular}
%    \label{eq:ordinalP}
%\end{equation}
%With $\sum_{k=1}^{K} \pi_{ij}(c_k|\theta_i) = 1$
Binary outcomes can be viewed as a special case of ordinal response with two categories.  Ordinal data assume that the categories are ordered ($c_1 < c_2 < \dots < c_K$), for instance as increasing levels of toxicity. It is natural to interpret models based on the cumulative response probabilities:

\begin{equation}
        \gamma_{ij}(c_k|\theta_i, t_{ij}) = Pr(Y_{ij}\leq c_k|\theta_i, t_{ij}) = \sum_{l=1}^k \pi_{ij}(c_l |\theta_i, t_{ij})  \text{   for k=1,\dots,(K-1)
        } 
    \label{eq:ordinalPC}
\end{equation}

%Equations \ref{eq:ordinalP} and \ref{eq:ordinalPC} are equivalent since it is possible to get $\pi_{ij}$ from $\gamma_{ij}$ :
%\begin{equation}
%    \begin{tabular}{lr}
%       $\pi_{ij}(c_k|\theta_i, t_{ij}) = \gamma_{ij}(c_{k+1}|\theta_i, t_{ij}) - \gamma_{ij}(c_k|\theta_i, t_{ij})$, & $k=1,\dots,K-1$.  \\
%    \end{tabular}
%\end{equation}

Classical models to describe ordinal data involve setting a parametric shape $f$ for a transformation of the probability $\gamma_{ij}$ using a link function.% $G$:
%\begin{equation}
% %        G(\gamma_{ij}(c_k|\theta_i, t_{ij})) = f(t_{ij},\theta_i, t_{ij}) 
%     \label{eq:ordinalmodel}
% \end{equation}

Commonly used link function are the logit, probit, log-log link and also the complementary log-log and cauchit link. $f$ is often modelled as a linear regression involving a time trend and potentially covariate predictors. The most popular model for ordinal data is the proportional odds model as presented in \cite{molenberghs_analyzing_2004} with a logit model for the link function and a linear function of $f$, yielding the following form:% for equation %\ref{eq:ordinalmodel}:
\begin{equation}
        logit(\gamma_{ij}(c_k|\theta_i, t_{ij})) = \sum_{k=1}^K \alpha_{ik} + \beta_i \cdot x(t_{ij})
    \label{eq:logitlinear}
\end{equation}
 where $x(t_{ij})$ is a vector of regression variables and $\beta_i$ a vector of coefficients.

\subsection{npd for ordinal data}

\subsubsection{Model evaluation for continuous responses}

Prediction discrepancies are defined as the quantile of each observation in its marginal predictive distribution by \cite{mentre_prediction_2006}. In NLMEM, we know how to write the probability of an observation conditionally to the individual parameters $\theta_i$ only, so the predictive distribution is obtained by integrating over the distribution of the random effects:
\begin{equation}
p_i(Y_{ij}|\psi, t_{ij}) = \int p(Y_{ij}|\theta_i, t_{ij},\psi)p(\theta_i|\psi, t_{ij})d\theta_i  \label{eq:density}
\end{equation}
When the structural model is nonlinear, there is no analytical solution for the integral in~(\ref{eq:density}).
$F_{ij}$ denotes the cumulative predictive distribution function of $Y_{ij}$ under the tested model. The prediction discrepancy $pd_{ij}$ for $Y_{ij}$ is defined as the value of $F_{ij}$ at the observation $Y_{ij}$:
\begin{equation}
pd_{ij} = F_{ij}(Y_{ij}|\psi) = \int^{Y_{ij}} p_i(y|\psi, t_{ij})dy = \int^{Y_{ij}} \int p(y|\theta_i, t_{ij},\psi)p(\theta_i|\psi, t_{ij})d\theta_i dy
\label{eq_pd_cont}
\end{equation}

By construction, the distribution of the $\pd$ are expected to follow a uniform distribution under $H_0$, however, their joint distribution in an individual needs to take into account their correlation.

\subsubsection{Extension of $\npd$ for categorical responses} 

By considering the $l^{th}$ modality ($l$ in ${1,\dots,K}$), $F_{ij}(c_l)$ is defined by:
\begin{equation}
F_{ij}(c_l|\psi) = \sum_{k=1}^{l} \pi_{ij}(c_k|\psi, t_{ij}) = \sum_{k=1}^{l} \int \pi_{ij}(c_k|\theta_i, t_{ij},\psi)p(\theta_i|\psi, t_{ij})d\theta_i 
\label{eq:F_cat}
\end{equation}

If all the subjects had the same design and covariates, we could consider the joint distribution of their discrete $\pd$ as defined by equation~\ref{eq:F_cat} and compare it to the expected distribution under the model. However, in practice the design variables and the number of observations may vary across subjects, so we propose to transform $\pd$ into a uniform variable that is easier to aggregate across the population. We consider the discrete values as interval-censored data from a uniform distribution, and we sample $\pd$ as previously proposed in \cite{cerou_development_2018, nguyen_extension_2012} for censored data as: 

\begin{equation}
\left \lbrace
\begin{tabular}{rlr}
     $pd_{ij}$ &$\sim~\mathcal{U}\left(F_{ij}\left(c_{l-1}\right) , F_{ij}\left(c_{l}\right)\right)$, & if $l>1$ \\
     $pd_{ij}$ &$\sim~\mathcal{U}\left(0 , F_{ij}\left(c_{l}\right)\right)$, & if $l=1$ \\
     $pd_{ij}$ &$\sim~\mathcal{U}\left(F_{ij}\left(c_{l-1}\right)  , 1\right)$, & if $l=K$ \\
\end{tabular}
\right.
\label{eq:pd_cat}
\end{equation}

If the model is correct (under $H_0$), i.e. the different values $F_{ij}(c_l)$ are well defined, then by construction $\pd$ follows a uniform distribution~$\mathcal{U}(0,1)$. Otherwise, some interval would be over/under-represented and the uniformity assumption violated.

Finally, we can also take the inverse function of the cumulative distribution function $\phi$ of $\mathcal{N}(0,1)$ to transform the $\pd$ to a normal distribution as many residual graphs use normally distributed variables.:
%\begin{equation}
%npd_{ij} = \phi^{-1} (pd_{ij})
%\label{eq:npde}
%\end{equation}

\subsubsection{Correction of the test statistic}

Under the assumption of independence, we can test the overall distribution of the $\npd$. \cite{brendel_evaluation_2010} suggest to use a combined test combining a normality, variance and mean test, or we can use an omnibus test such as the Kolmogorov-Smirnov test. To account for the correlation between observations within individuals, which increases the type I error of that test, we propose to estimate the distribution of the test statistic for the Kolmogorov-Smirnov test between the distribution of the $\npd$ and a $\mathcal{N}(0,1)$ distribution by using $B$ simulations under $H_0$. A threshold corresponding to the 95$^{\rm th}$ percentile of the distribution of this test statistics is computed and is used to correct the power.

\subsection{Chi-square test}

An alternative test can be used when all subjects are followed at the exact same times ($t_j (1 \leq j \leq n)$) and no covariate enters the model. In this case, all subjects have the same predictive distributions at each time, and we can test for model adequacy using a Chi-square test. %At time $j$, assuming the number of observations for outcome $k$ is $O_{j,k}= \sum_{i=1}^{N} 1\!\!1_{Y_{ij} = c_k}$, we can compare with the expected number $E_{j,k}$ of observations for that outcome under the model through the Chi-squared statistic: 
%\begin{equation}
%    \chi^2_j = %\sum_{k=1}^{K}\frac%{\left( O_{j,k} - %E_{j,k} %\right)^2}{E_{j,k}} %\label{eq:statChi2}
%\end{equation}
We use the asymptotic approximation and compare the value of $\chi^2_j$ to $\chi^2_{K-1}$, a $\chi^2$ distribution with $K-1$ degrees of freedom. 

For longitudinal data, we test the $n\times p$ statistics for each level of stratification ($n$ visits and across all other possible $p$ combination of covariates), and we correct $\alpha_{j}$ using a Bonferroni correction to ensure an overall level of $\alpha=0.05$ ($\alpha_j=\frac{\alpha}{n\times p}$). To do so, we approximate by simulation the Chi-square statistics on each level of stratification and estimate for each one the $\alpha_j$ threshold. Then we compute the Chi-square statistic when comparing the model and the data and the model is rejected if at least one of the Chi-square statistic $\chi_j^2$ is lower than their associated approximated threshold. 

%$E_{j,k}$ is here computed using simulations as:
%\begin{equation}
%    E_{j,k} = \frac{1}{R} \sum_{i=1}^{N}\sum_{r=1}^{R} 1\!\!1_{Y_{ij}^{sim(r)} = c_k}
%\end{equation}

\subsection{Evaluating $\npd$ performance for binary data in a simulation study}

\subsubsection{Simulation model}

We evaluated the performance of the test based on the npd in the context of binary data. This example is inspired from the work of ~\cite{Seurat}, who proposed optimal designs to estimate the parameters of a linear logistic model describing a binomial response, in the presence of a treatment effect $trt$ ($trt=1$ under treatment, 0 otherwise): 

% Eco: corrigé l'équation pour Y=1
%    logit\left(P(Y \ge k)\right) = \theta_1 + \left(\theta_2 + \beta \times trt  \right)t
\begin{equation}
    logit\left(P(Y=1)\right) = \theta_1 + \left(\theta_2 + \beta \times trt  \right)t
    \label{eq:M1}
\end{equation}
where $\theta_1$ represents the intercept, $\theta_2$ the slope and $\beta$ the treatment effect. Parameters $\theta_x$ were assumed to follow a normal distribution: $\theta_x = \mu_x + b_x$ where $b_x \sim N(0,\omega_x^2)$ for $x=\{1,2\}$. We assumed 4 measurement times at $t=\{0,2,11,12\}$ months.

\subsubsection{Simulations scenarios}

We define one scenario as one combination of a true model $M_B$ and a tested model $M_V$. A model is defined by the parameter's value and the structural shape of the model. In each scenario, $B=200$ datasets are generated under the model $M_B$, and $V=1000$ simulations are generated under the model $M_V$. In each scenario, datasets were successively generated with three sample sizes ($N=\{50, 100, 274\}$) subjects with the same design ($50\%$ in each treatment group).

The base definition of the model $M_B$ is parameter estimates defined in 
Table~1, with eq~\ref{eq:M1} describing the structural shape of the model.
%Table~\ref{tab:ex1}, with eq~\ref{eq:M1} describing the structural shape of the model.

We considered two sets of simulations, a first set with the same structural model changing the value of one parameter, and a second set evaluating different model shapes. 

\paragraph*{Misspecification of parameter values}

Data were generated according to the base scenario but with different values of parameters: $\mu_1 = \{-4,3,-2,1,0\}$, $\mu_2 = \{-0.3, -0.09,\\0, 0.09, 0.3\}$, $\beta = \{0,0.3,0.45,0.7,1\}$, $\omega_1 = \{0.17,0.3,0.5,0.7,1\}$ and $\omega_2 = \{0.1, 0.17,0.3,\\0.5, 0.7\}$. Since we considered one parameter change at a time, this resulted in 25 combinations of parameters used to generate the data. For each parameter of interest and for each change, the 5 parameter values used to generate the data were used in the model to evaluate ($M_V$). This resulted in 125 scenarios ($25$ models $M_B$ and for each 5 models $M_V$), including 25 models where the tested model was the true model ($M_V=M_B$).

\paragraph*{Misspecification of the structural model}

We used the same design and settings as in the first scenarios, but considered 4 different models: linear model (M1) described in eq. \ref{eq:M1}, loglinear model (M2), quadratic model (M3) and exponential model (M4), represented by the equations:
%{\it Data were generated according to the 4 following models: linear model (M1, \ref{eq:M1}), loglinear model (M2), quadratic model (M3) and exponential model (M4).}

    \begin{eqnarray}
        M2 : logit\left(P(Y \ge k)\right) & = & \theta_1 + \left(\theta_2 + \beta \times trt  \right)\log(t+1)  \\
        M3 : logit\left(P(Y \ge k)\right) & = & \theta_1 + \left(\theta_2 + \beta \times trt  \right)t^2  \\
        M4 : logit\left(P(Y \ge k)\right) & = & \theta_1 + \left(\theta_2 + \beta \times trt  \right)\left(\exp(\theta_3 t)-1 \right)
    \end{eqnarray}

    % \begin{tabular}{ccl}
    %     M2 :& $logit\left(P(Y \ge k)\right)$ =& $\theta_1 + \left(\theta_2 + \beta \times trt  \right)\log(t+1)$  \\
    %     M3 :& $logit\left(P(Y \ge k)\right)$ =& $\theta_1 + \left(\theta_2 + \beta \times trt  \right)t^2$  \\
    %     M4 :& $logit\left(P(Y \ge k)\right)$ =& $\theta_1 + \left(\theta_2 + \beta \times trt  \right)\left(\exp(\theta_3 t)-1 \right)$  \\
    % \end{tabular}

Parameter values for M2 to M4 were chosen in order to have the same mean value of the logit of the probability as in M1 at baseline (t=0 month) and at the end (t=12 month) of the study in the two treatment groups. Parameter values are presented in 
Table~2. 
%Table~\ref{tab:ex1_mis2}. 

As previously, for each scenario, one model was used in $M_B$ and the 4 models are successively used as $M_V$, and for each 4 models. This resulted in 16 scenarios, including 4 where $M_V=M_B$. 

\subsubsection{Evaluation}

The scenarios where the same model was used to simulate the data and compute the $\npd$ were used to build the reference distribution for the test based on the $\npd$ (simulations under $H_0$) and compute the threshold to ensure a p-value of 5\%. This threshold was then used to correct the power for the test in the 4 scenarios with different parameter values (simulations under $H_1$). The power was computed as the fraction of simulations where the test based on $\npd$ rejected the null hypothesis.

We compared the test based on $\npd$ to the Chi-square test described above, by accounting on two levels of stratification, visit and treatment group, with the Bonferroni correction applied to account on the 8~tests (4 recording time and 2 treatments).

We used the statistical software \cite{rcoreteam} version 3.5.1 for all the simulations, computations and graphs in this study.

\subsection{Application to toenail data}

\hskip 18pt To illustrate the use of $\npd$ applied to real data, we considered binary data from a randomised clinical trial comparing two treatments for fungal toenail infection, and available in {\sf R} as the {\sf toenail} dataset in the {\sf R} package "prLogistic". 

Data are from \cite{debacker_toenail}, a multi-center randomised comparison  of two oral treatments (A and B) for toenail infection. 294 patients are measured at seven visits, i.e. at baseline (week 0), and at weeks 4, 8, 12, 24, 36, and 48 thereafter, comprising a total of 1908 measurements. The primary end point was the absence of toenail infection and the outcome of interest is the binary variable "onycholysis" which indicates the degree of separation of the nail plate from the nail-bed (none or mild versus moderate or severe).

Several analyses have been made, like in %\citep{lesaffre2001effect,lin2011goodness,fitzmaurice2012applied,tan
\cite{lesaffre2001effect,lin2011goodness} and the logistic random effect model developed by \cite{hedeker1994random} is considered in this work. This model includes a random intercept ($\beta_1$ and $\omega_1$), time ($\beta_2$) and treatment (A or B) ($\beta_3$) as covariate. We considered the interaction term between time and treatment but no treatment effect alone as it would impact the intercept which shouldn't be different between arms due to the randomisation process. 

The logistic random effect model is fitted using \cite{Monolix2019} software, which uses SAEM algorithm for population parameter estimation and importance sampling to compute the log likelihood estimation. We performed 1000 Monte-Carlo simulations through the associated R package "mlxR", in order to compute the $\npd$ and its associated test. 

\section{Results} \label{sec:results}

\subsection{Evaluating npd performance}

We first present the results of the simulation study with binary data. In the following, we present directly the corrected power in all scenarios, using the threshold for each test statistic computed under H0. Note that the resulting type I error equals exactly 0.05 with the npd test, while there was an inherent variability for the Chi-square test. Indeed, for the Chi-square test a Bonferroni correction is applied to each of the 8 tests (one for each visit and treatment group) using the threshold computed for the overall test statistic, which induces variability especially in the datasets with a small sample size.

\subsubsection{Misspecification on parameter values}

\begin{figure}[!h]
    \centering
    \includegraphics[width=0.85\linewidth]{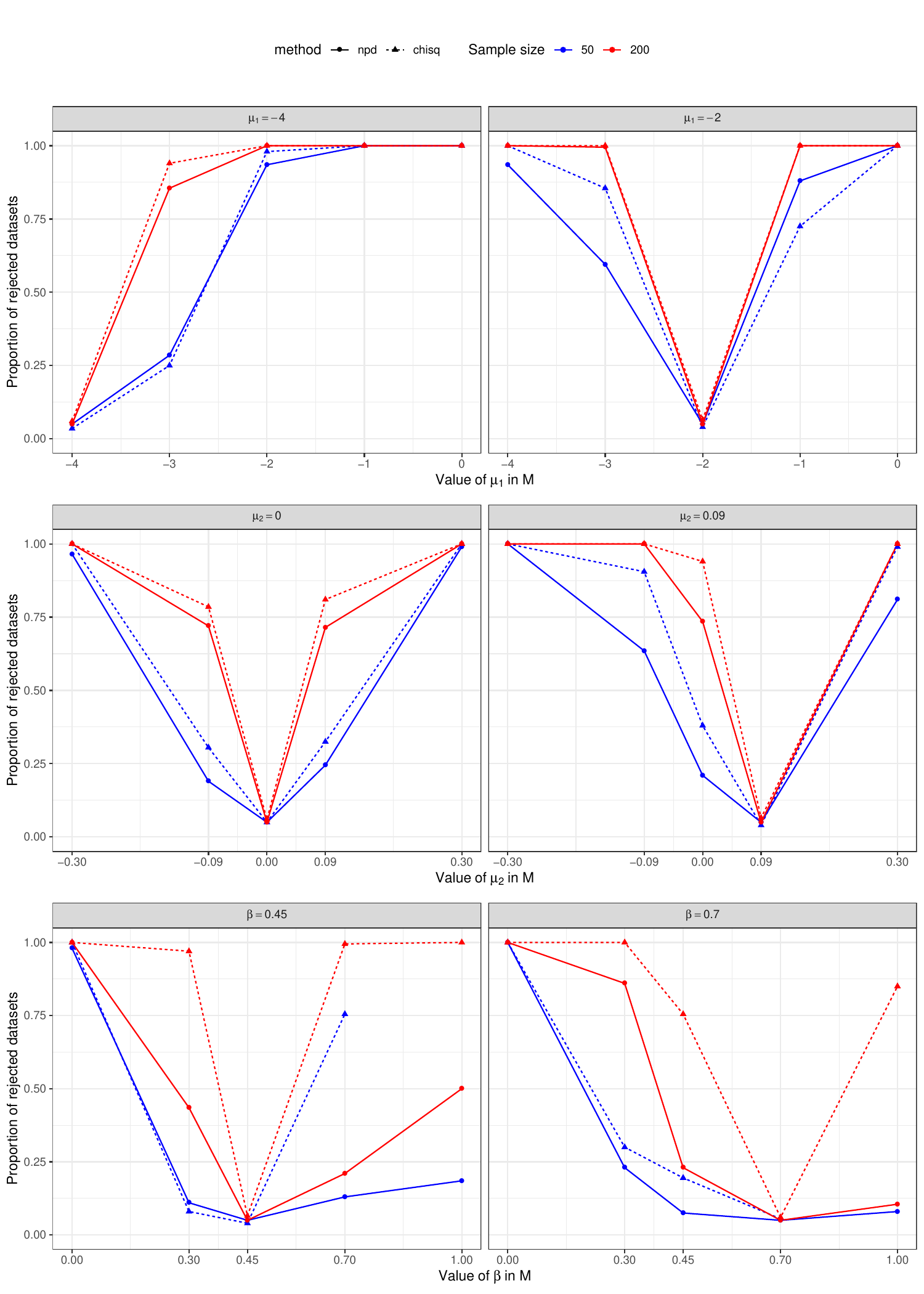}
    \caption{Power of the npd compared to a Chi-square test in case of parameter's misspecification on the fixed effect level, depending on three sample size}
    \label{fig:power_theta}
\end{figure}

\begin{figure}[!h]
    \centering
    \includegraphics[width=0.9\linewidth]{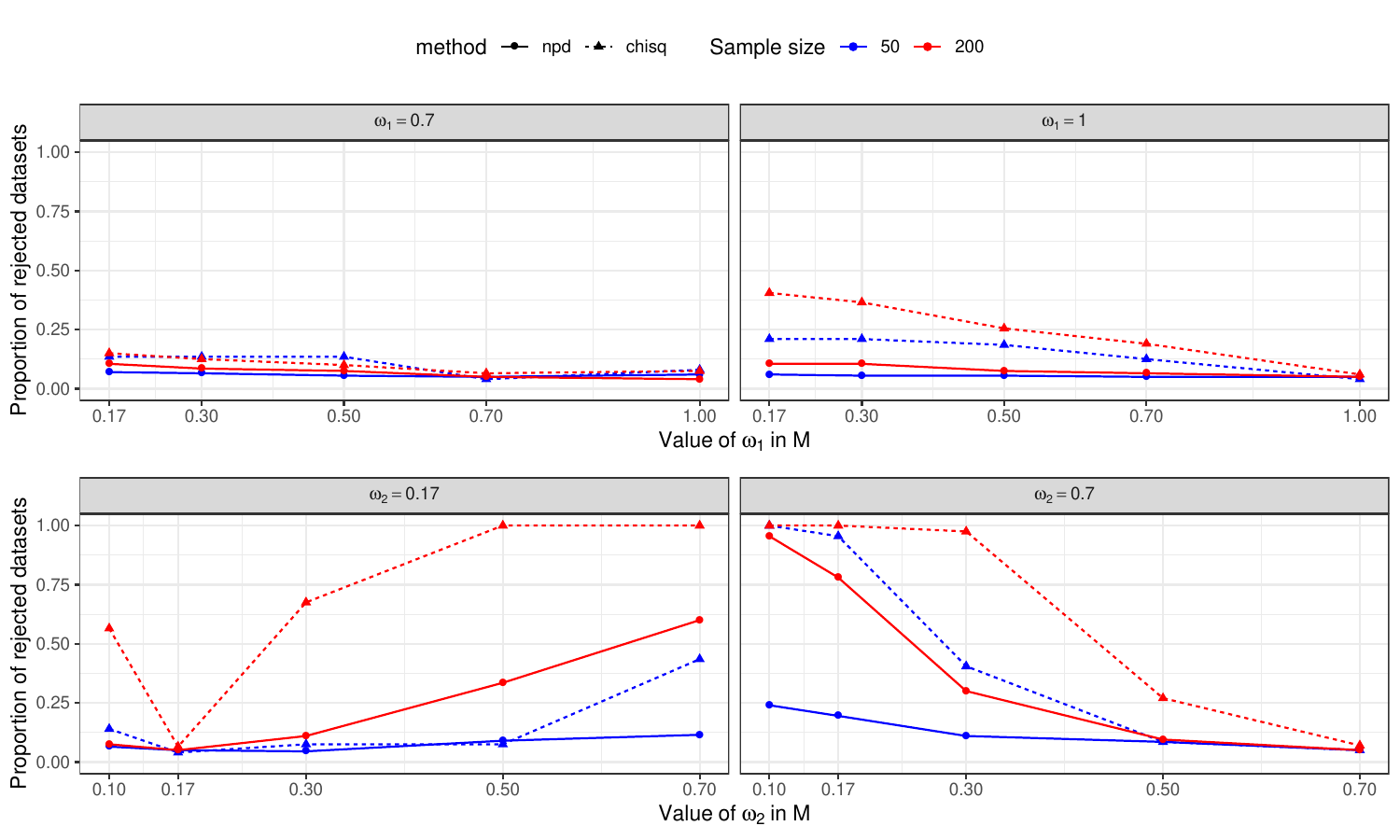}
    \caption{Power of the npd compared to a Chi-square test in case of parameter's misspecification, on the between subject variability level, depending on three sample size}
    \label{fig:power_omega}
\end{figure}

In the first set of simulations, we studied the effect of misspecification on the parameter values by moving the value of one parameter away from the simulated values. %For each fixed, treatment or random effect, parameter we considered two simulation values and four tested models (the true model and three models with misspecified values of the parameters) 
and we show the results in Fig.~\ref{fig:power_theta} for fixed effects and Fig.~\ref{fig:power_omega} for random effects. In each figure, the different parameters investigated are shown on a separate row of plots, and the lines represent the corrected power of the test based on npd (solid line and circles) and of the Chi-square test of proportion (dashed lines and triangles), with different colours identifying the sample size. Simulations were performed also for N=100 but were omitted from the plots for more clarity, as the results were always intermediate between the N=50 and N=200 groups, as expected.

Fig.~\ref{fig:power_theta} shows the corrected power for the intercept $\mu_1$ (top row) time effect $\mu_2$ (middle row) and treatment effect $\beta$ (bottom row). The change in power has the expected pattern in each plot, with an increase according to the rise in the size of the population and an increased difference from the true value, used to generate the data. In all the scenarios, the power is systematically slightly higher for the Chi-square test. The power to detect a model misspecification was generally high.

Fig.~\ref{fig:power_omega} shows the corrected power to detect a misspecification on the between subject variability of intercept and time effect parameter (respectively $\omega_1$ and $\omega_2$). As previously, the power increases with a rise in the sample size and an increased difference from the true value. However, the power is overall quite small, especially when the value of $\omega_2$ used to generate the data is 0.3 (less than 30\%). The power is even smaller with a misspecification on $\omega_1$ where both tests failed to detect of model misspecification 70\% to 95\% of the time, even with a large number of subjects. 

\subsubsection{Misspecification on the structural model}

\begin{figure}[!h]
    \centering
    \includegraphics[width=0.9\linewidth]{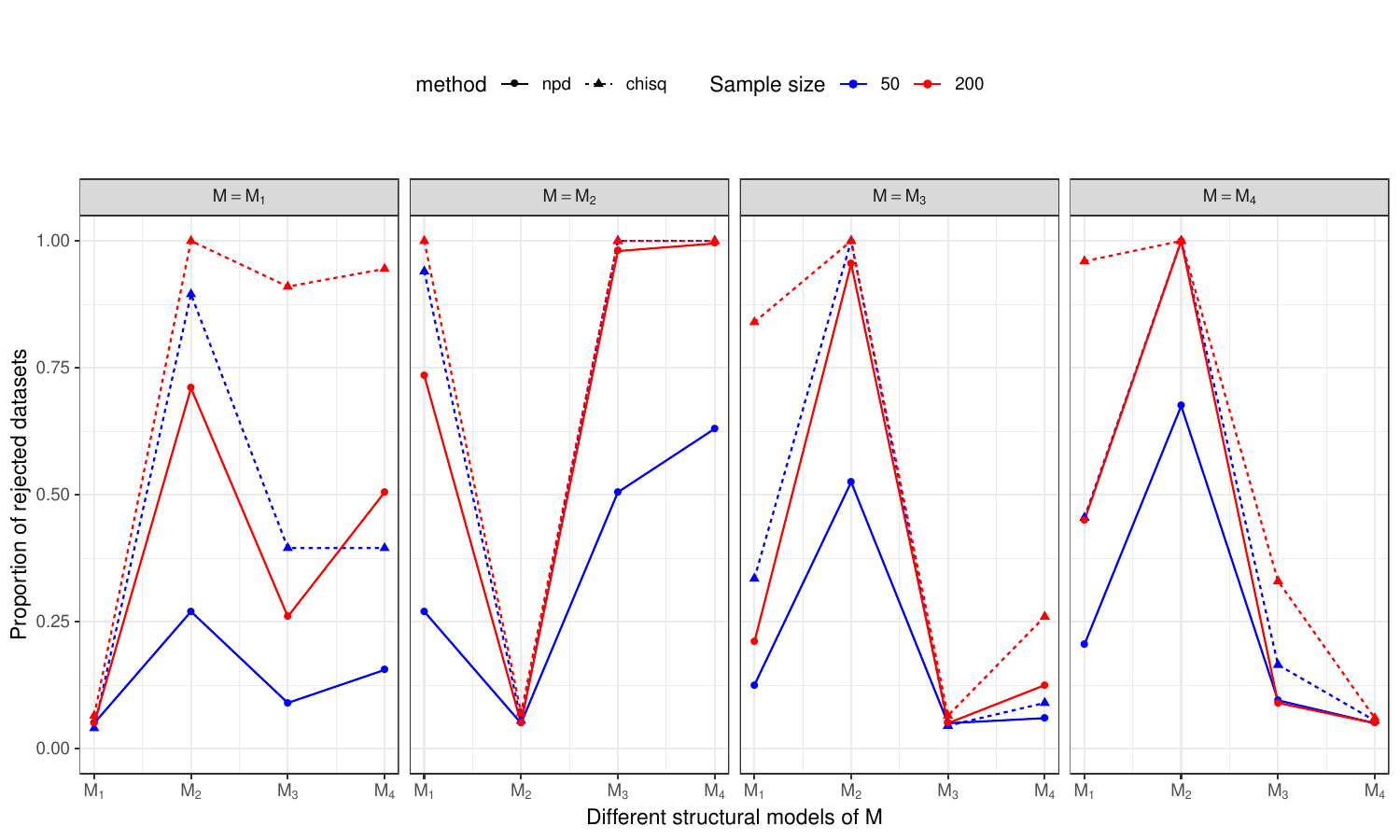}
    \includegraphics[width=0.9\linewidth]{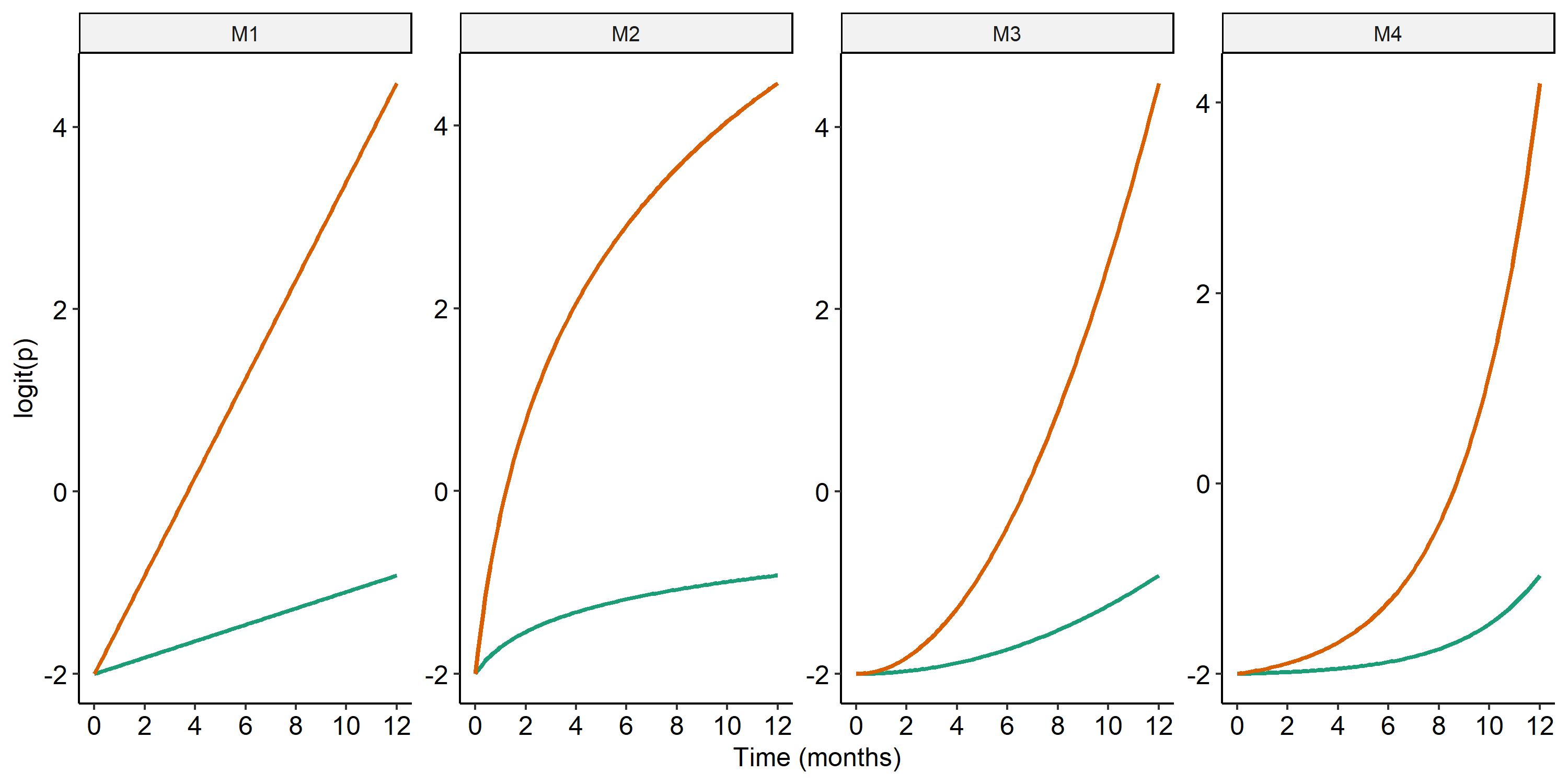}
    \caption{Power of the npd compared to a Chi-square test in case of misspecification on the structural model, for three sample sizes.}
    \label{fig:power_model}
\end{figure}

\begin{figure}[!h]
    \centering
    \includegraphics[width=0.9\linewidth]{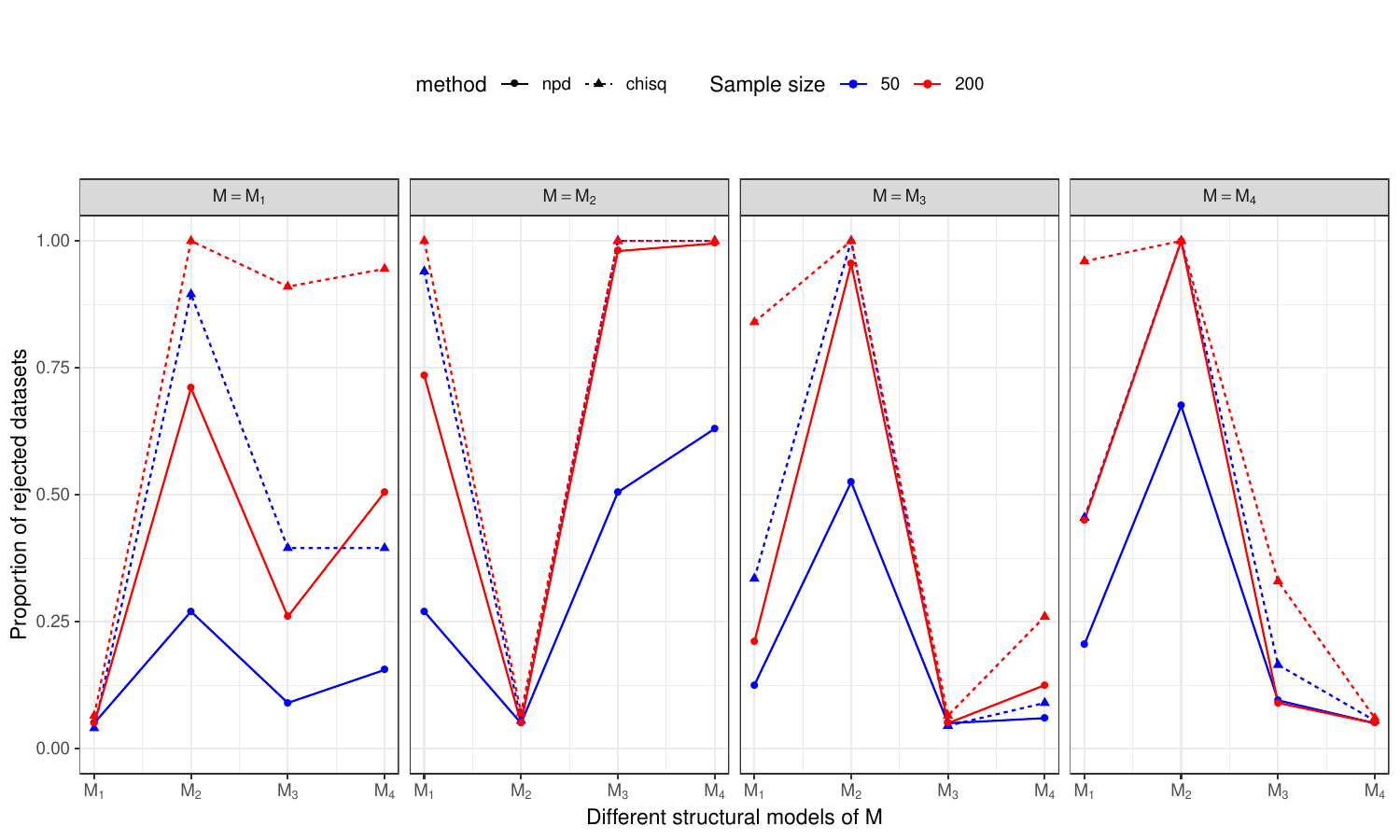}
    \includegraphics[width=0.9\linewidth]{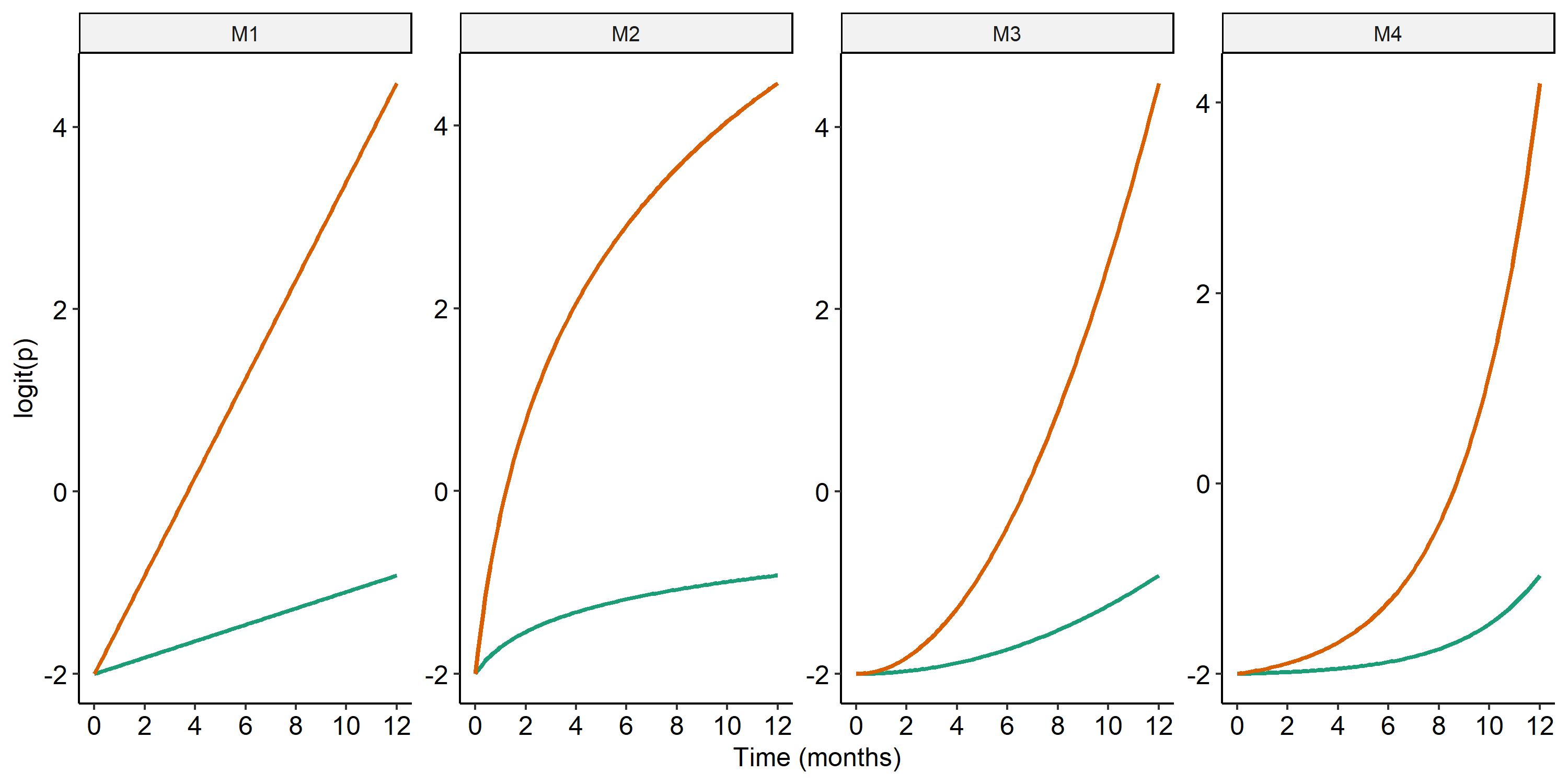}
    \caption{Prediction from the four models of the logit probability of response over time, for both the control and treated group.}
    \label{fig:model_profile}
\end{figure}

In a second series of simulations, we investigated misspecifications in the structural model. The results are shown in Fig.~\ref{fig:power_model} for four models. In each cell, we simulate under one model and compare to four alternatives including the true one. Compared to the previous results, we do not expect an increased power as model differs in the "x-axis", as there is no order in those models. Instead, the difference between those models should be evaluated in term of difference in the shape of the response. Fig.~\ref{fig:model_profile} represents the patterns of predicted logit-probabilities over time for the same models, to illustrate their evolution with time and the impact of the expected treatment effect. 

As previously, the power increases with sample size. According to those settings, the power to detect a difference between the quadratic ($M_3$) and the exponential model ($M_4$) is small with the Chi-square test. With the test based on $\npd$, it is not possible to detect a difference as the power did not differ from $5\%$. The log-logistic model ($M_2$) is very different in this context to the others. Indeed there is a high power to detect that this model is not used to generate the data ($M = \{M_1,M_3,M_4\}$) or to detect that other models are not used to generate the data ($M=M_2$). 
In some settings (eg comparing M1 to M4), the increase in power for the Chi-square test was up to 70\% compared to that of the corrected test based on $\npd$, while it was closer to 10\% when comparing M1 and M3. The bottom plots show that the power increases as the shapes of the model differ, and the Chi-square test may be more sensitive to a large difference at one visit or between the two treatment groups.

\subsection{Stratifying the npd test over visits and treatment}  \label{sec:strat_test}

In all the scenarios presented above (misspecifications on parameter values and structural model), we also computed the npd test based on each level of stratification (visits:treatment) and a Bonferroni correction to ensure an overall threshold of 5\%. Figures in supplementary material S1, S2, and S3
% \ref{fig:power_theta_strat}, \ref{fig:power_omega_strat}, and \ref{fig:power_model_strat} 
shows that stratifying on visits and treatment results in a power increase, without however reaching the power level of the Chi-square test in all scenarios, excepted for the scenarios with parameter misspecifications on $\mu_1$ and $\mu_2$ where there is a power decrease.

\subsection{Application to toenail data}

\begin{figure}[!h]
    \centering %encore pas finie 
    \includegraphics[width=0.6\linewidth]{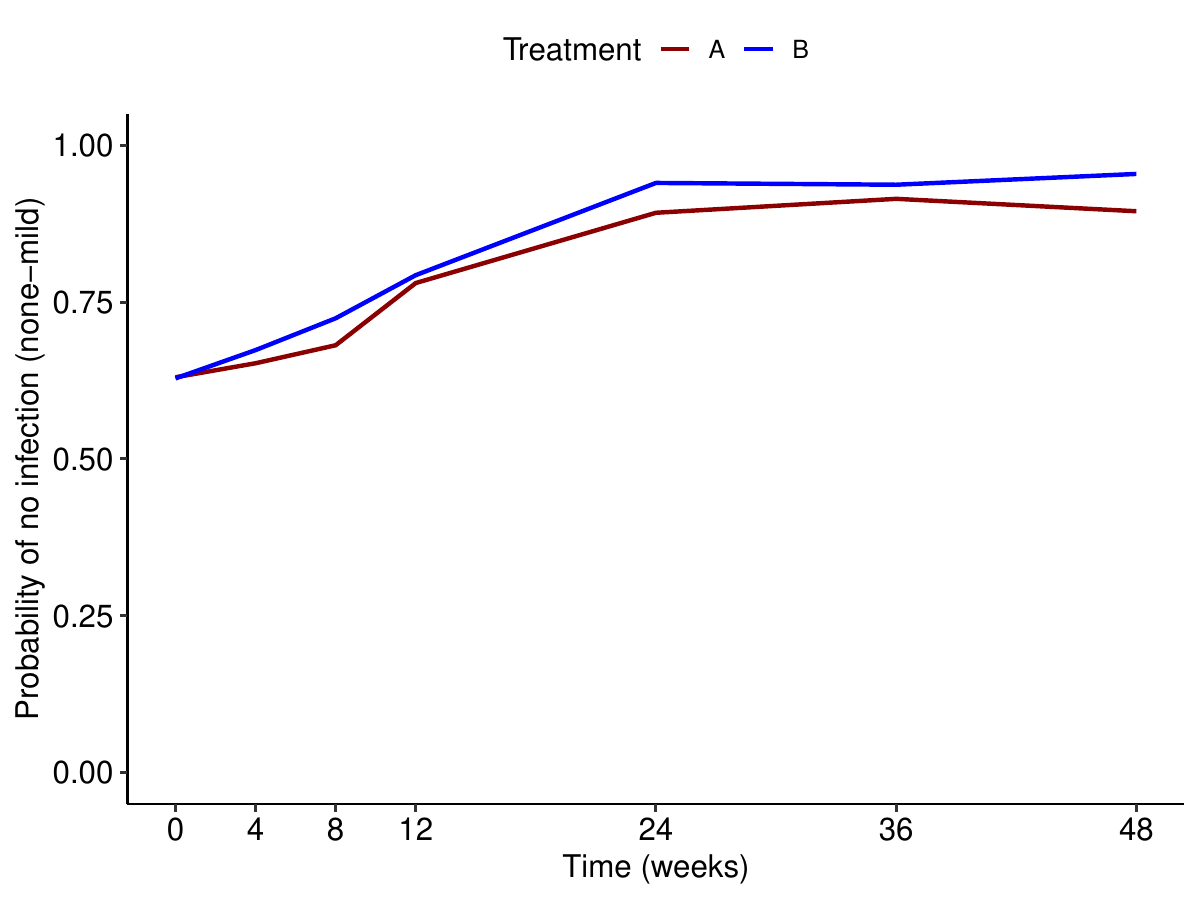}
    \caption{Probability of being non infected over time (Y=0 if moderate or severe infection, Y=1 if mild or absent infection), stratified by treatment arm. }
    \label{fig:toenail_data}
\end{figure}

One of the clinical outcome of interest in this study was the clearance of infection, so in this analysis we model the outcome "none or mild degree of separation of the nail plate from the nail-bed", with values of 1 for subjects presenting mild or no infection, and 0 for subjects with moderate of severe infection. Fig.~\ref{fig:toenail_data} shows that the probability of being infection-free increases over time in both treatment groups, with a slight difference between the two treatments A and B. 

\subsubsection{Model building}

We tested several models to describe the probability of no infection ($P(Y=1)$), representing different assumptions on time trends and treatment effect:
\begin{eqnarray}
    M_{\text{constant,no trt}} : logit\left(P(Y=1)\right) & = & \theta_1 \\ 
    M_{\text{linear,no trt}} : logit\left(P(Y=1)\right) & = & \theta_1 + \theta_2 t \\ 
    M_{\text{linear,trt:$\theta_1$}} : logit\left(P(Y=1)\right) & = & \left(\theta_1 + \beta \times trt \right) + \theta_2 t \\
    M_{\text{linear,trt:$\theta_2$}} : logit\left(P(Y=1)\right) & = & \theta_1 + \left(\theta_2 + \beta \times trt  \right)t
\end{eqnarray}

The parameters of these models are $\theta_1$, the intercept of the linear model on the logit of the probability, $\theta_2$, the linear coefficient describing the time effect, and $\beta$, the treatment effect for treatment B (treatment A is taken as the reference in the model), which could be applied to either $\theta_1$ or $\theta_2$. Other models to describe the effect of time such as loglinear or quadratic models were also tested but the parameters could not be estimated satisfactorily and we did not investigate them further. 

%Table~\ref{tab:comparisonFit} 
Table~3 
shows the estimated statistical criterion (equal to minus twice the log-likelihood) for each of the tested models. As they are all nested within another, the p-value of the Likelihood ratio test is given relative to the simpler model to which they are compared, and we chose a significance threshold of 5\% to select models.  As shown in the table, introducing a time parameter led to a very significant drop in the likelihood. We then tested the effect of treatment on the baseline probability of no infection ($M_{\text{linear,trt:$\theta_1$}}$), which was not significant, consistent with the randomisation process. On the other hand, there was a slightly significant decrease in the likelihood when assuming different slopes according to the treatment group, with treatment B slightly more effective than treatment A. The final model was therefore $M_{\text{linear,trt:$\theta_2$}}$.

The parameter estimates of the best model are given in Table~\ref{tab:fit_estimates}. All parameters were well estimated with low relative standard error (RSE). The Wald test associated to the treatment effect on slope parameter was significant ($p < 0.003$). 

The probability of being infection-free at baseline was estimated to be 0.65, with no difference between the two groups. At the end of the study, this probability had increased to 0.92 in treatment A and 0.97 in treatment B.

\subsubsection{Model evaluation using npd}

\begin{figure}[!h]
    \centering
    \includegraphics[width=1\linewidth]{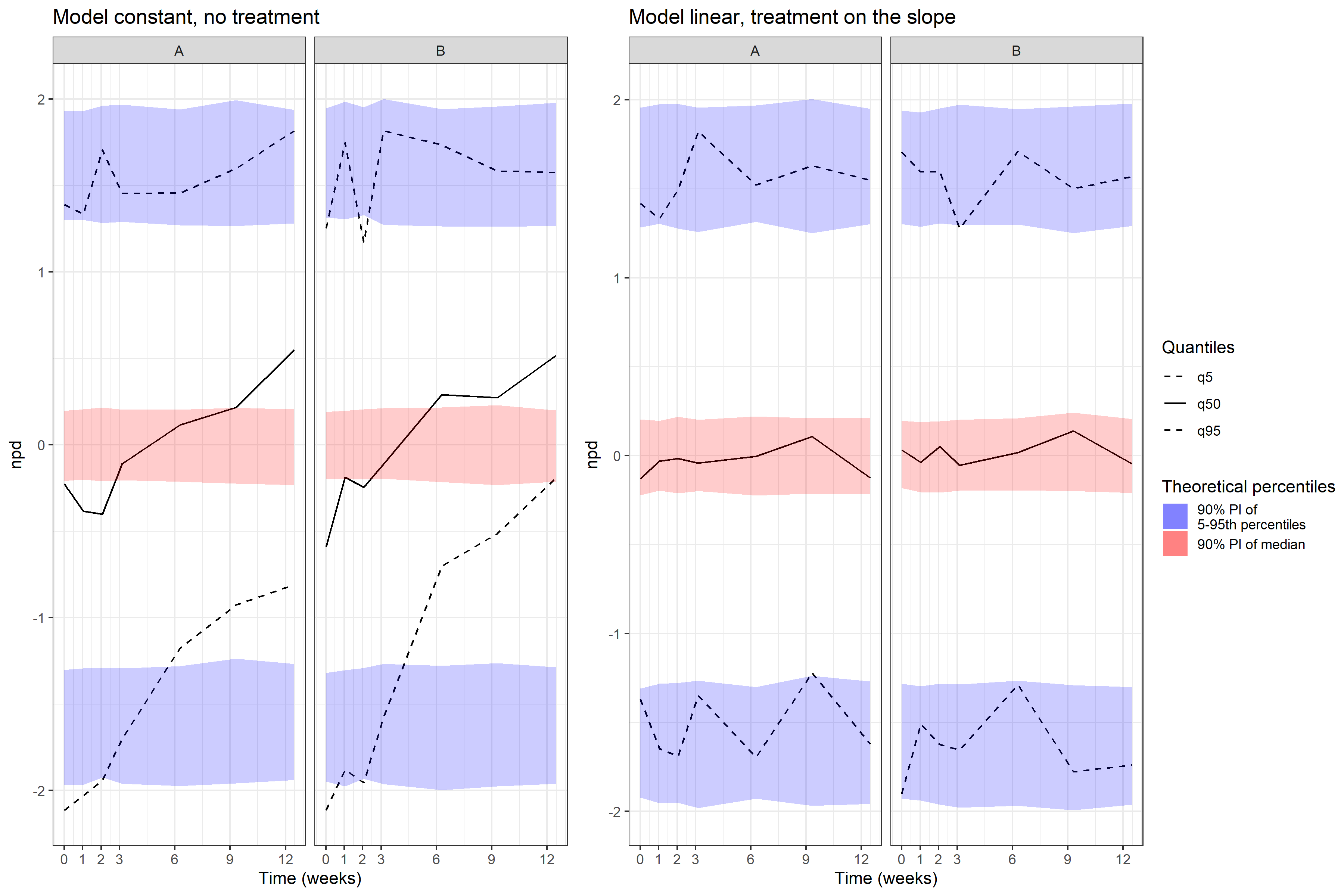}
    \caption{Scatter plot of the npd versus time, stratified by treatment group (left: base model model$M_{\text{constant,no trt}}$; right: final model $M_{\text{linear,trt:$\theta_2$}}$).}
    \label{fig:scatter}
\end{figure}

Fig.~\ref{fig:scatter} shows diagnostic graphs using $\npd$ for base model $M_{\text{constant,no trt}}$ (left) and final model $M_{\text{linear,trt:$\theta_2$}}$ (right). As for continuous data (see e.g. ~\cite{comets_model_2010, nguyen_model_2017}), these graphs compare observed percentiles of the $\npd$ (the median, as a solid line, and the 5$^{\rm th}$ and 95$^{\rm th}$ percentiles, in dashed lines) at each time point to prediction intervals, in pink for the median and in blue for the extreme percentiles. These prediction intervals are obtained by simulating from the theoretical normal distribution at each time point and computing the 95\% prediction interval of each observed percentile. With continuous data we usually overlay the observed $\npd$, but with the jittering involved in creating them for categorical data the individual $\npd$ do not have the usual interpretation as quantiles and we prefer summarising them using the observed percentiles for clarity. We show the diagnostic graphs stratified by treatment group, but the same trends are seen in the overall plots (not shown).

On the left, a clear trend can be seen for both the median and the 95$^{\rm th}$ percentile of the observed $\npd$, when compared to the corresponding prediction intervals, in both treatment groups, with $\npd$ decreasing from mostly positive to mostly negative values over time. This corresponds to an overprediction of the probability of no infection at the beginning of the study, evolving in a compensatory underprediction at late times. It suggests that the change in risk over time should therefore be included in the model. A linear (on the logit-scale) time effect has been included in the model on the right, as well as a treatment effect, and we see that now the observed percentiles of the $\npd$ remain in their respective prediction intervals, suggesting no further model misspecification. Of note, similar diagnostic graphs as those on the right of Fig.~\ref{fig:scatter} were obtained with the second best model ($M_{\text{linear,no trt}}$), indicating these diagnostic graphs could not distinguish between the models with or without treatment effect. This is consistent with the small difference in likelihood seen in 
Table~3, which suggests only a small statistical advantage in adding the treatment effect to the slope. 
%Table~\ref{tab:comparisonFit}, which suggests only a small statistical advantage in adding the treatment effect to the slope. 

\begin{figure}[!h]
    \centering
    \includegraphics[width=0.5\linewidth]{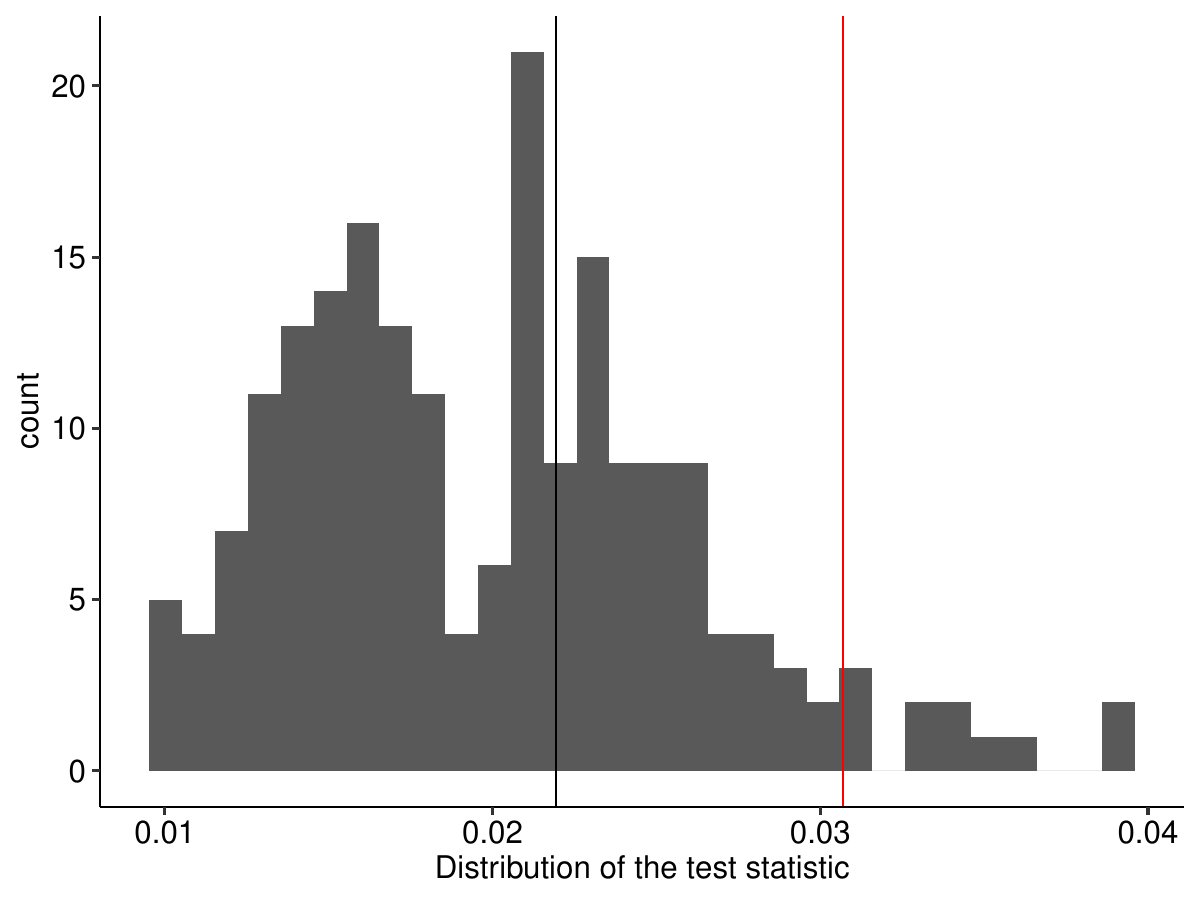}
    \caption{Distribution of the test statistic under the null hypothesis, with the threshold associated with the 95$^{\rm th}$ percentile (red), compared to the statistics based on the data.}
    \label{fig:distrib_stat}
\end{figure}

In parallel to these graphs, we can compute the p-value for the test based on $\npd$. Fig.~\ref{fig:distrib_stat} illustrates the distribution of the test statistic under the null hypothesis as a histogram: the red vertical line indicates the 95$^{\rm th}$ percentile of the distribution under the null, while the black vertical line indicates the value computed under model $M_{\text{linear,trt:$\theta_2$}}$. The graph indicates that the observed value is compatible with the distribution under the null, so that we do not reject the hypothesis that this model adequately describes the data. In line with the previous result, the corresponding statistic for the model without treatment effect ($M_{\text{linear,no trt}}$) was also not significant.

\section{Discussion} \label{sec:discussion}

%General  
\hskip 18pt In the present manuscript, we present an extension of $\npd$ to mixed effect models involving discrete data models with categorical data. We assess their performance in a simulation study, and illustrate their use through an analysis of binary repeated data in the presence of a treatment effect. \cite{nguyen_model_2017} stated that $\npd$ and $\npde$ are part of the tools recommended to evaluate non-linear mixed effect models describing continuous data, and ~\cite{cerou_development_2018} have recently proposed extensions to these diagnostics for survival data. 

% Estimation and evaluation
As summarised by \cite{lavielle_mixed_2014}, models for repeated discrete data are defined in terms of probability, through the likelihood of observing a given outcome, in contrast to continuous data which are directly observed. A natural evaluation criterion with continuous data is to consider the difference between the observed value and the prediction of the model, called a residual, but this doesn't readily translate to categorical data as the observations and the model predictions are not in the same scale. Attempts have been made to define residuals in logistic regression for ordinal data, such as the Pearson, cumulative Pearson and deviance residuals in the work of \cite{mccullaghGLM}. Pearson residuals aggregate data from the same category to compare the observed proportion of events to the expectation predicted by the model. Deviance residuals measure the contribution to the log-likelihood and indicates poorly fitted values. These residuals however are difficult to assess, because while their asymptotic distribution is a $\chi^2$, they have no known distribution in small samples, with a mean that can be different from 0. As synthesised in  \cite{liu_residuals_2018}, residuals based on the sum of cumulative residuals were developed but implicitly assume an equal distance between ordinal categories% \cite{liu_graphical_2009}. Li and Sheperd proposed 
Sign-based statistic (SBS) residuals were proposed but rely mainly on the zero mean under the null hypothesis property.% \cite{li_new_2012}. Liu and Zhang \cite{liu_residuals_2018} 
They recently proposed surrogate residuals, the idea being to define a continuous variable S as a “surrogate” of Y and then obtain residuals based on S. They proposed a specific computation for general models which rely on a jittering method, either on the outcome scale, either on the probability scale. They showed that those surrogate residuals have good properties with respect to mean structures, link functions, heteroscedasticity, proportionality. However, their properties and extension to non-linear mixed effect models have not been studied.

%% Our finding
% Remplacer par une phrase
%The approach we present in this manuscript is an extension of the $\npd$, developed previously for non-linear mixed effect models with continuous data, to categorical data. $\npd$ are simulation based residuals defined for an observation as the quantile of this observation within its own marginal predictive distribution. Using each observation's own predictive distribution ensures the resulting residuals take into account differences in design variables as well as the non-linearity in the model~\cite{brendel_metrics_2006}. With categorical data however, because we model the probability but observe only discrete events, quantiles are limited to a discrete set. We therefore proposed to use a jittering method, a method similar to that used with the "surrogate" residuals presented by Liu and Zhang~\cite{liu_residuals_2018} in order to transform the quantiles to a uniform distribution, then to a normal distribution. With a similar interpretation as in the continuous test, we can derive a statistical test for the $\npd$. The type I error of the test is inflated because of the correlation between the observations collected in the same subjects~\cite{mentre_prediction_2006}, and the decorrelation method proposed for continuous data by~\cite{brendel_metrics_2006} cannot be applied. Thus, we proposed to approximate the distribution of $\npd$ under the null hypothesis to correct the test. 

Defining residuals as quantiles from the cumulative prediction distribution was proposed for logistic and linear regression by ~\cite{Dunn96}. Our $\npd$ can be viewed as an extension of this approach for mixed effect models, and we derive the distribution of the test statistic by simulations to account for the correlation between observations. We evaluated the performance of this test in a simple scenario with a homogeneous design composed of two treatment groups observed at the same four times. We found that $\npd$ showed good power to detect different types of model misspecification including different model shapes for the evolution of the probability of outcome with time, or changes in the fixed effect parameters, but had low power to detect misspecification in the random effect parameters.

In this simple setting, we could compare the performance of the test based on the $\npd$ with the performance of the Chi-square test, testing whether the proportion of events observed in the simulated dataset corresponded to the expected proportion. The Chi-square test is only valid when all the variables have the same distribution, so we needed to perform separate tests for different visits and different treatment groups, and combine the separate tests, here with a Bonferroni correction to ensure the global test remains at the 5\% threshold. The Chi-square test was generally more powerful than the $\npd$ over a range of alternatives, but we observed similar trends in power across the different simulation scenarios with both tests. Here, the Chi-square test could be considered here as a gold standard given the large number of observations and the fact that only 2 categories were simulated. Its good performance in our simulation study is in part due to large differences in some combinations of visit/treatment where the Chi-square test was very powerful even after a Bonferroni correction which led to an overall rejection of the null hypothesis. The Chi-square test can be generalised to more than two categories through a Chi-square test, and we expect that the power for the $\npd$ will become closer to that of the Chi-square test as the number of categories increase. However, the Chi-square test crucially depends on being able to define a single proportion of events at each time point for each treatment group, while the $\npd$ adjust to the circumstances of each observation through its predictive distribution, and can thus be applied regardless of the design and model without stratification. Another advantage of the $\npd$ is that it can be used to assess the model graphically, as we illustrated in the analysis of the toenail data, while the Chi-square test only assesses aggregate data. 

Finally, the figures in supplementary material shows an increase in the power of the test stratified on the covariates. This can be explained by the balancing of the $\npd$ distributions between the different levels of stratification when the test is global, which could lead to a non-rejection of the test. The stratified test thus allows to highlight the differences in distribution but as the power is not systematically higher, we suggest to evaluate the stratified $\npd$ distribution to evaluate covariate effects graphically, as recommended by \cite{brendel_evaluation_2010} for continuous data. 

% Toenail
For the toenail data, the Chi-square test could not be applied as the number of observations was different at the various visits, illustrating the difficulty of applying a Chi-square test in the presence of unbalanced design or many covariates.

The toenail dataset was chosen to illustrate the $\npd$ on a real case study because it is freely available on the R package "prLogistic", and it has been extensively modelled in the literature with reasonably simple models used to characterise the data. %The response of interest in this work is a dichotomised onchyolosis endpoint (degree of separation of the nail plate from the nail-bed: none and mild versus moderate and severe). 
This data has been widely used to showcase analyses and algorithms for generalised mixed effect models. When the repeated data aspect was investigated, they have been most often described by a logistic random-effect models including a random intercept representing patient variability at baseline. Several models have been presented to describe the drug effect and \citep{lin2011goodness} presented this effect either on slope, either both the intercept and the slope. %, which have been presented either on the intercept \citep{lesaffre2001effect}, either on the slope, either on both parameters \citep{lin2011goodness}. 
We found that the model which best describe this data is a model with a treatment effect on slope. The treatment effect was only slightly significant, however, which, along with the fact that some authors preferred to test a treatment effect on the intercept, may explain the different models found in the literature. Our results are consistent with \cite{lin2011goodness} with similar parameter estimates and parameter precision. Although we could show the importance of taking into account the evolution of infection probability with time, models with or without treatment effect had similar diagnostic plots, in line with values of the log-likelihood remaining in close range for these models. 

In the analysis of the toenail data, we ignored missing data, implicitly assuming that the missing data was missing at random or completely at random. In our analysis, we consider models similar to those used in the analyses reported in~\cite{verbeke1997linear}, which did not consider dropout, as our focus was more on model evaluation. It is worth noting however that only 76\% of the subjects in the toenail study completed the seven scheduled visits, and one question is whether this is an acceptable assumption. \cite{verbeke2000model} considered this issue in their analysis of the continuous data associated to the toenail data, combining a marginal model with a quadratic evolution for the unaffected nail length at each visit with a logistic regression model for dropout. They find some evidence that dropout from the study is not completely random. We could therefore consider jointly modelling the dropout from the study along with the probability of response. 
%However, they also note that the results are highly sensitive to model assumptions, and in particular to the random effect structure, and their analysis of the continuous data relies on a quadratic model and no other alternative models were tested. 
%In another chapter of the book, they suggested missing data may be associated with patterns of missingness with patients not showing up at intermediate visits, rather than dropout from the study.

In this example, the $\npd$ proved very versatile as a model evaluation tool, providing compelling diagnostic graphs across a variety of models with a similar interpretation as the diagnostic plots obtained for continuous data. One of the major strengths of npd is that they can be visualised despite complex design, through the quantile-quantile plot, over time, continuous variable, and stratified by the covariates.  
%Another simulation-based diagnostic graphs with widespread application are the VPC, and VPC for categorical data compare the observed proportion of each category of event to a prediction interval. 
%For example, in the context of IRT, Ueckert showed those VPC for two of the rheumatoid arthritis score items. In such graphs for the overall evaluation, there is as many levels of stratification as the number of level and items. This is de-multiplied by adding other levels of stratification according to levels of covariates (an example in the IRT context and in supplementary material of Lyauk work \cite{lyauk2020item} with 168 graphs needed to evaluate the model, i.e. 6 scores possible on 7 items, and 4 dose levels). 

Of course, the test based on $\npd$ has limits. It proved generally less powerful than a stratified Chi-square test in our simulations, suggesting the transformation of the discrete quantile distribution to a continuous distribution by the jittering of the quantiles within uniform intervals incurs a loss of power which may be quite severe in some scenarios. It should be noted however that the simulation study involved a binary variable, so that the informativeness of the data was very limited, and only 4 visits, so that the Bonferroni correction applied to the Chi-square test was not too severe. We surmise that the gap should narrow when applied to variables with more categories, such as score data from cognitive decline studies, or count data. Another point to note is that, as with all simulation-based metrics, simulations must be feasible which may be problematic (or even impossible), for example in the context of adaptive dose regimens, or patients who switch between treatments because of recorded/not recorded outcome. Another limit is the need to compute the distribution of the test statistics by simulation to correct for the inflation due to repeated observations within a subject. A practical issue is the number of replicates used to build this distribution, which needs to be large enough to correctly approximate the p-value. In this work, the number of replicates used to estimate the distribution of the test statistic was 200 replicates. This provided a good performance of the corrected test but it could of interest to evaluate the performance of the test by increasing the number of replicates to build the distribution of the test statistic. A better alternative to building the distribution under the null hypothesis would be to transform the vector of $\pd$ or $\npd$ to ensure independence, as was proposed for continuous data in ~\cite{brendel_metrics_2006}. For discrete data, we could consider copulas to describe the dependence between random variables. Although they have been widely used in quantitative finance or in joint model with longitudinal measurements and survival times such as in the work of \cite{ganjali2015copula}, their application to vectors of quantiles of unequal size with varying dependency structure appears at the moment challenging.

\section{Conclusion}

In this article, we formally extend $\npd$ to discrete data, by transforming discrete quantiles to normal variables to produce diagnostic graphs. We showed the good performance of the test based these $\npd$ in a simulation study modelling the probability of no infection as a function of treatment and visits, complementing their use as visual diagnostics.
The extension to nominal data is straightforward by assigning an artificial "order" between categories. For graphical evaluation, specific adjustments should be implemented for meaningful diagnostics. Finally, a natural extension would be to apply the same idea to count data, which has a similar nature but a potentially infinite number of categories. It could be interesting to verify the $\npd$ properties in this context, and further evaluations will focus on models for count data and score data with multiple categories.

%%%%%%%%%%%%%%%%%%%%%%%%%%%%%%%%%%%%%%%%%%%%%%
%% Funding information, if any,             %%
%% should be provided in the                %%
%% funding section.                         %%
%%%%%%%%%%%%%%%%%%%%%%%%%%%%%%%%%%%%%%%%%%%%%%
\section*{Acknowledgements}

{\bf Funding:} Marc Cerou received funding from Institut de Recherches Internationales Servier for this work, as part of a PhD research fellowship programme. Code implementing the npde for models with binary or categorical data is available on request.

%%%%%%%%%%%%%%%%%%%%%%%%%%%%%%%%%%%%%%%%%%%%%%
%% Supplementary Material, including data   %%
%% sets and code, should be provided in     %%
%% {supplement} environment with title      %%
%% and short description. It cannot be      %%
%% available exclusively as external link.  %%
%% All Supplementary Material must be       %%
%% available to the reader on Project       %%
%% Euclid with the published article.       %%
%%%%%%%%%%%%%%%%%%%%%%%%%%%%%%%%%%%%%%%%%%%%%%

\clearpage
\newpage
\renewcommand\refname{References}
% Submissions are not required to reflect the precise reference formatting of the journal (use of italics, bold etc.), however it is important that all key elements of each reference are included.
\bibliographystyle{plainnat}
\bibliography{biblio}

\newpage

\section*{Figures \& Tables}

%\phantom{Invisible text to adjust section label and Fig. and Tab.}

\clearpage
\begin{table}
    \caption{Parameter values for the logistic regression model on binomial data in~\cite{Seurat}. } \label{tab:ex1}

    \centering
    \begin{tabular}{cccc}
        \hline
        Parameter & Fixed-effect & Transformation & Between subject standard deviation \\
        \hline
        $\mu_1$ & -2 & Normal & 0.7\\
        $\mu_2$ & 0.09 & Normal & 0.17 \\
        $\beta$ & 0.45 & - & -\\
        \hline
    \end{tabular}
\end{table}

\begin{table}
     \caption{Parameter values for each structural model. M1 is the linear model, M2 the loglinear model, M3 the quadratic model, M4 the exponential model. } \label{tab:ex1_mis2}
    \centering
    \begin{tabular}{ccccc}
        \hline
        Parameter & M1 & M2 & M3 & M4 \\
        \hline
        $\mu_1$ & -2 & -2 & -2& -2\\
        $\mu_2$ & 0.09 & 0.42 & $7.50 \times 10^{-3}$ & $2.01 \times 10^{-2}$\\
        $\mu_3$ & - & - & - & 0.33\\
        $\beta$ & 0.4500 & 2.100 & 0.0375& 0.1005 \\
        $\omega_1$ & 0.7 & 0.7 & 0.7& 0.7\\
        $\omega_2$ & 0.17 & 0.79 & $1.41 \times 10^{-2}$ & $3.79 \times 10^{-2}$\\
        \hline
    \end{tabular}
\end{table}

\begin{table}
    \caption{Model selection process for the toenail data. The p-value from the log-likelihood rat\-io test is given in the last column, with the parent model for the test.} 
    \label{tab:comparisonFit}
    \centering
    \begin{tabular}{cccc}
         \hline
         Model & -2LL & Model compared & p-value \\
         \hline
         $M_{\text{constant,no trt}}$ & 2737.7 & -& - \\
         $M_{\text{linear,no trt}}$ & 1257.0 & $M_{\text{constant,no trt}}$ & p$<$0.001 \\
         $M_{\text{linear,trt:$\theta_1$}}$ & 1255.3  & $M_{\text{linear,no trt}}$& NS\\
         $M_{\text{linear,trt:$\theta_2$}}$ & 1251.9  & $M_{\text{linear,no trt}}$ & p=0.041\\
         \hline
    \end{tabular}
\end{table}

\begin{table}
    \caption{Parameter estimates for the best model ($M_{\text{linear,trt:$\theta_2$}}$) adjusted to the toenail data, along with their standard error of estimation (SE) and relative SE (RSE). \label{tab:fit_estimates}}
    \centering

    \begin{tabular}{cccc}
         \hline
         Parameter & Value & SE & RSE (\%)\\
         \hline
         $\theta_1$ & 1.76 & 0.330 & 19 \\
         $\theta_2$ & 0.36 & 0.039 & 11 \\
         $\beta$ & 0.19 & 0.064 & 33\\
         $\omega_1$ & 4.05 & 0.370 & 9\\
         \hline
    \end{tabular}
\end{table}

% \section{Supplementary Material}

% Supplementary material for this paper is provided as an additional PDF file.

% \newpage
% \beginsupplement

% % \setcounter{figure}{0}

% \begin{figure}[!h]
%     \centering
%     %\includegraphics[width=0.9\linewidth]{power_theta_strat_time_trt.pdf}
%     %\caption{Power of the npd with a test stratified by time and treatment, compared to a Chi-square test in case of parameter's misspecification on the fixed effect level, depending on three sample size}
%     \label{fig:power_theta_strat}
% \end{figure}

% \begin{figure}[!h]
%     \centering
%     %\includegraphics[width=0.9\linewidth]{power_omega_strat_time_trt.pdf}
%     %\caption{Power of the npd with a test stratified by time and treatment, compared to a Chi-square test in case of parameter's misspecification, on the between subject variability level, depending on three sample size}
%     \label{fig:power_omega_strat}
% \end{figure}

% \begin{figure}[!h]
%     \centering
%     %\includegraphics[width=0.9\linewidth]{power_model_strat_time_trt.pdf}
%     %\caption{Power of the npd with a test stratified by time and treatment, compared to a Chi-square test in case of misspecification on the structural model, depending on three sample size.}    \label{fig:power_model_strat}
% \end{figure}

\end{document}